\documentstyle[twoside,fleqn,npb,epsf]{article}
\def\lsim{\mathrel{\rlap{\lower4pt\hbox{\hskip1pt$\sim$}}
    \raise1pt\hbox{$<$}}}                % less than or approx. symbol
\def\gsim{\mathrel{\rlap{\lower4pt\hbox{\hskip1pt$\sim$}}
    \raise1pt\hbox{$>$}}}                % greater than or approx. symbol

\def\3{\ss}

\newcommand{\AmS}{{\protect\the\textfont2
  A\kern-.1667em\lower.5ex\hbox{M}\kern-.125emS}}

\title{
       \vspace{-3.65cm}                                     % for preprint
       {\normalsize DESY 99--064}    \\[-0.2cm]             % for preprint
       {\normalsize HUB--EP--99/24}  \\[-0.2cm]             % for preprint
       {\normalsize FUB-HEP/1-99}    \\[-0.2cm]             % for preprint
       {\normalsize May 1999}        \\                     % for preprint
        \vspace{1.78cm}                                     % for 3 preprint #
        Towards a lattice calculation of $\Delta q$ and $\delta q$
            \thanks{Talk given by R. Horsley at DIS99,      % for preprint
                    DESY-Zeuthen.}}                         % for preprint

\author{S.~Capitani%
           \address{Deutsches Elektronen-Synchrotron DESY,
                    D-22603 Hamburg, Germany},
        M.~G\"ockeler%
           \address{Institut f\"ur Theoretische Physik, Universit\"at
                    Regensburg, D-93040 Regensburg, Germany},
        R.~Horsley%
           \address{Institut f\"ur Physik, Humboldt-Universit\"at zu Berlin,
                    D-10115 Berlin, Germany},
        H.~Perlt%
           \address{Institut f\"ur Theoretische Physik, Universit\"at
                    Leipzig, D-04109 Leipzig, Germany},
        D.~Petters%
           \address{Deutsches Elektronen-Synchrotron DESY and NIC,
                    D-15735 Zeuthen, Germany}
                    \hspace{-0.25cm} $^{,}$ \hspace{-0.2cm}
           \address{Institut f\"ur Theoretische Physik,
                    Freie Universit\"at Berlin, D-14195 Berlin, Germany},
        D.~Pleiter$^{\rm e,}$ \hspace{-0.2cm} $^{\rm f}$,
        P.~E.~L. Rakow$^{\rm b}$,
        G.~Schierholz$^{\rm a,}$ \hspace{-0.2cm} $^{\rm e}$,
        A.~Schiller$^{\rm d}$
        and
        P.~Stephenson%
           \address{Dipartimento di Fisica,
                    Universit\`a degli Studi di Pisa e INFN,
                    Sezione di Pisa, 56100 Pisa, Italy}}

\begin{document}

\begin{abstract}
Within the framework of lattice $QCD$ a high statistics computation
of the nucleon axial and tensor charges is given. Particular attention is
paid to the chiral and continuum extrapolations.
\end{abstract}

\maketitle

\section{Introduction}
\label{introduction}

The axial and tensor charges of the nucleon are defined by
($s^2 = - m_N^2$):
\begin{eqnarray}
   \langle \vec{p}, \vec{s}| {\cal A}^{\mu} | \vec{p}, \vec{s} \rangle
       &=& 2 s^\mu \Delta q(\mu)
                                                           \\
   \langle \vec{p}, \vec{s}|{\cal T}^{\mu\nu} | \vec{p}, \vec{s} \rangle
       &=& {2 \over m_N} \left( s^\mu p^\nu -
                                s^\nu p^\mu \right) \delta q(\mu)
                                                \nonumber
\label{tensor_charges}
\end{eqnarray}
with ${\cal A}^\mu = \overline{q} \gamma^\mu \gamma_5 q$,
${\cal T}^{\mu\nu} = i \overline{q} \sigma^{\mu\nu} \gamma_5 q$.
The chirality and charge conjugation of
these operators are $(+,+)$ and $(-,-)$ respectively.
The charges have a parton model interpretation:
\begin{eqnarray}
   \Delta q(\mu) &=& \int_0^1 dx \left[
       \{q_\uparrow(x,\mu)-q_\downarrow(x,\mu)\} +
                                 \right.        \nonumber  \\
                 & & \qquad \left.
       \{\overline{q}_\uparrow(x,\mu)-\overline{q}_\downarrow(x,\mu)\} \right]
                                                \nonumber  \\
   \delta q(\mu) &=& \int_0^1 dx \left[
       \{q_\bot(x,\mu)-q_\top(x,\mu)\} -
                                 \right.        \nonumber  \\
                 & & \qquad \left.
       \{\overline{q}_\bot(x,\mu)-\overline{q}_\top(x,\mu)\} \right]
\label{parton}
\end{eqnarray}
where $x$ is the fraction of nucleon momentum carried by parton
quark density $q_{\bullet}(x,\mu)$ at scale $\mu$, in scheme ${\cal S}$.
$\Delta q$ is related to the lowest moment of the $g_1$ structure function
and can be measured in (polarised) DIS, while $h_1$
(for $\delta q$) having a $-$ chirality must be found
in a reaction allowing the quarks to have a different chirality,
such as (polarised) Drell-Yan.

Known for the charges is that $\Delta u - \Delta d = g_A \approx 1.26$,
(from neutron decay) and in the heavy quark limit $\Delta u = \delta u = 4/3$,
$\Delta d = \delta d = -1/3$. Indeed in the non-relativistic limit
as fermions (ie here quarks) are eigenstates of $\gamma_0$ then
we expect $\Delta q = \delta q$.

\section{The Lattice Approach}
\label{lattice}

The hypercubic discretisation of the Euclidean path integral for $QCD$ with
lattice spacing $a$ and resulting Monte Carlo evaluation of the
partition function allows, in principle, a fundamental test of $QCD$.
That being said, the lattice programme is rather like an experiment:
careful account must be taken of error estimations and extrapolations.
There are three limits to consider:
\begin{enumerate}
   \item The box size must be large enough so that finite
         size effects are small. Currently sizes of $\sim 2\mbox{fm}$
         seem large enough (the $N$ diameter is $\sim0.8\mbox{fm}$).
   \item The chiral limit, when the quark mass approaches zero.
         It is difficult to calculate quark propagators at quark masses
         much below the strange quark mass. For each $a$ value,
         three (or four) heavier quark masses are used and a linear
         extrapolation is made to the chiral limit, see
         Fig.~\ref{fig_ga_aor02_p0_DIS99_joint}.
         \begin{figure}[t]
            \vspace*{-0.75cm}
            \epsfxsize=7.00cm \epsfbox{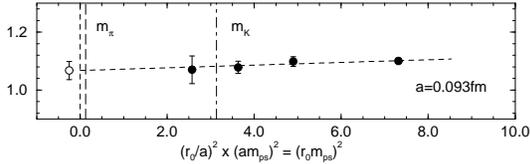}
            \vspace*{-1.00cm}
            \caption{\footnotesize The $\chi$-extrapolation for $g_A$.}
            \vspace*{-0.75cm}
            \label{fig_ga_aor02_p0_DIS99_joint}
         \end{figure}
         (Within our precision, there is no difference to extrapolating
         to the $u/d$ quark mass or to the chiral limit.)
   \item The continuum limit, $a^2 \to 0$ (the leading order
         discretisation effects of the action can be so arranged
         to be $O(a^2)$, \cite{jansen95a}). We have performed
         simulations at three $a$ values corresponding approximately to
         (0.093, 0.068, 0.051) $\mbox{fm}$,
         [$a^{-1} =$ (2.12, 2.90, 3.85)$\mbox{GeV}$]
         using scale $r_0 = 0.5\mbox{fm}$, [$r_0^{-1} = 395\mbox{MeV}$],
         \cite{guagnelli98a}.
\end{enumerate}
Our lattice calculation of matrix elements is standard,
see eg \cite{gockeler95a}. We only note here that we are working in
the `quenched approximation' when the fermion determinant in the partition
function is set equal to one. The nucleon matrix elements consist of 
two diagrams: a quark line connected piece and a quark line
disconnected piece ($qldis$).
Only the quark line connected piece is calculated.
Thus, strictly speaking, from eq.~(\ref{parton}), we can only compute
$\delta u - \delta d$ and $\Delta u - \Delta d$. However, 
due to the additional negative sign in eq.~(\ref{parton}) for $\delta q$,
we might also expect to be able to determine $\delta u$, $\delta d$
separately. (Both the above described approximations allow
considerable savings in computer time.)

\section{Renormalisation}
\label{renormalisation}

\begin{figure}[h]
   \vspace*{-0.50cm}
   \epsfxsize=6.50cm \epsfbox{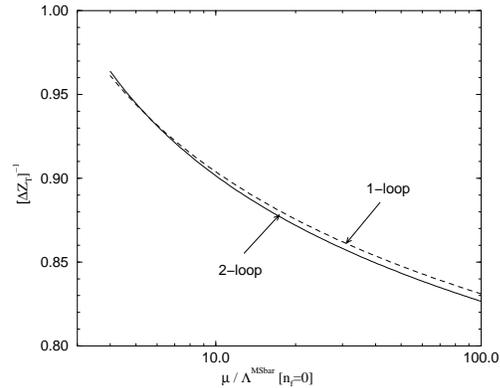}
   \vspace*{-1.00cm}
   \caption{\footnotesize
            $[\Delta Z_{\cal T}^{\overline{MS}}]^{-1}$ using the one
            and two loop expansions of the beta and gamma functions,
            \cite{artru90a} in units of $\Lambda^{\overline{MS}}$
            for $n_f=0$ flavours. The relation to
            the scale $r_0$ is given in
            \cite{capitani98a}. So, eg, at $\mu=1$, $2$, $5$ $\mbox{GeV}$,
            $[\Delta Z_{\cal T}^{\overline{MS}}]^{-1}=$ $0.959^7_7$,
            $0.911^5_4$, $0.870^3_3$ respectively.}
   \vspace*{-0.75cm}
   \label{fig_hMsbar_o_mrgi_DIS99}
\end{figure}

\begin{figure}[h]
   \vspace*{-0.50cm}
   \epsfxsize=6.50cm \epsfbox{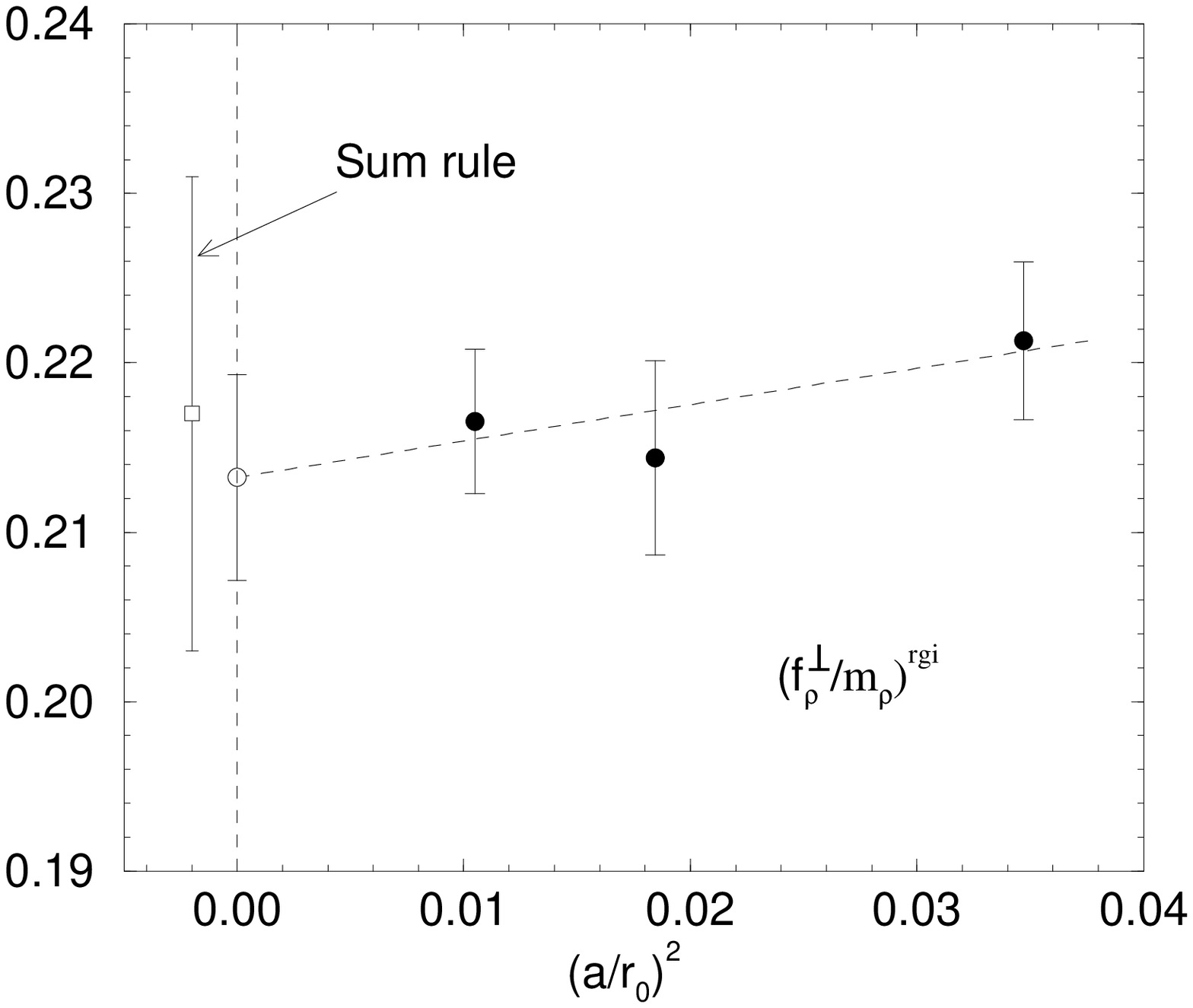}
   \vspace*{-1.00cm}
   \caption{\footnotesize
            The $a^2\to 0$ limit for $f_\rho^{\bot rgi} / m_\rho$.}
   \label{fig_f_rhoT_o_mrho_aor02_p0_DIS99}
   \vspace*{-0.75cm}
\end{figure}

In general operators (or matrix elements) must be renormalised
${\cal O}_R^{\cal S}(\mu)= Z_{\cal O}^{\cal S}(\mu) {\cal O}_{bare}$
(in scheme ${\cal S}$, at scale $\mu$) before they
can be compared with experimental results.
For the axial current ensuring $PCAC$ on the lattice gives $Z_{\cal A}$
and has been calculated in \cite{luscher96a}. The tensor current
is more complicated. It is first convenient to define a $rgi$--operator,
which is both scheme and scale independent by
\begin{equation}
   {\cal T}^{rgi} = \Delta Z_{\cal T}^{\cal S}(\mu) {\cal T}_R^{\cal S}(\mu)
\label{rgi_def}
\end{equation}
with
\begin{eqnarray}
   \lefteqn{\left[\Delta Z_{\cal T}^{\cal S}(\mu)\right]^{-1} =}
     &                                                \\
     &  \left[ 2b_0 (g^{\cal S})^2 \right]^{d_{{\cal T},0}\over 2b_0}
        \exp \left[ \int_0^{g^{\cal S}} \! \! \! d\xi
        \left[ {\gamma_{\cal T}^{\cal S}(\xi)
                          \over \beta^{\cal S}(\xi)} +
              {d_{{\cal T},0}\over b_0 \xi} \right] \right]
                                           \nonumber
\end{eqnarray}
The factor $[\Delta Z_{\cal T}^{\overline{MS}}]^{-1}$ is plotted in
Fig.~\ref{fig_hMsbar_o_mrgi_DIS99}.
Given ${\cal T}^{rgi}$ then from Fig.~\ref{fig_hMsbar_o_mrgi_DIS99}
and eq.~(\ref{rgi_def}) we can thus find ${\cal T}_R^{\overline{MS}}(\mu)$
at any desired scale $\mu$. For $Z_{\cal T}^{\overline{MS}}$
only the result for one loop perturbation theory is known.
A non-perturbative result would be desirable, at present
we shall use `tadpole improved' perturbation theory, which re-expands
the perturbation series, \cite{capitani97a},
using a physical coupling constant and 
removes the lattice `tadpole' diagrams. Our explicit procedure using
$g^{\overline{MS}}$ is described in \cite{gockeler97a}.
To check that this gives a plausible result, we
calculate the $\rho$-tensor decay constant, which also requires the same
renormalisation constant. This decay constant is defined by
\begin{equation}
   \langle 0| {\cal T}^{\mu\nu} | \rho(\vec{p}, \lambda) \rangle
       = i \left( e^\mu_\lambda p^\nu - e^\nu_\lambda p^\mu \right)
             f_\rho^{\bot}(\mu)
\end{equation}
This gives for $f_\rho^{\bot rgi} / m_\rho$ the results
%($a^2 \to 0$) shown in Fig.~\ref{fig_f_rhoT_o_mrho_aor02_p0_DIS99}.
shown in Fig.~\ref{fig_f_rhoT_o_mrho_aor02_p0_DIS99}.
Ball and Braun, \cite{ball96a}, using the sum rule approach find a value for
$f_\rho^{\bot}(\mu=1\mbox{GeV})$ of $160\pm10\mbox{MeV}$ ($n_f=3$).
Using Fig.~\ref{fig_hMsbar_o_mrgi_DIS99}, this converts to
$f_\rho^{\bot rgi} / m_\rho = 0.217(14)$.
From Fig.~\ref{fig_f_rhoT_o_mrho_aor02_p0_DIS99} we see that there is
good agreement between the lattice and sum rule methods.

\section{Results}
\label{section}

\begin{figure}[h]
   \vspace*{-0.50cm}
   \epsfxsize=6.50cm \epsfbox{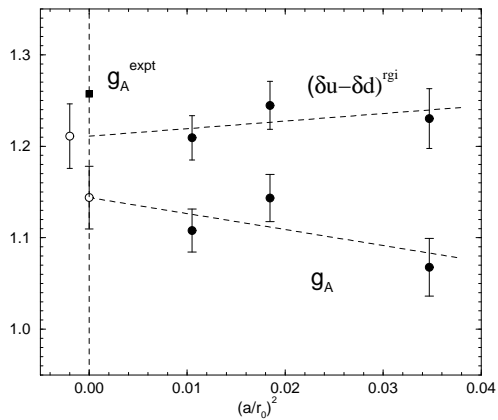}
   \vspace*{-1.00cm}
   \caption{\footnotesize
            The $a^2 \to 0$ limit for $g_A$ and
            $(\delta u - \delta d)^{rgi}$.
            Also shown is the experimental result for $g_A$.}
   \vspace*{-0.75cm}
   \label{fig_ga_aor02_p0_DIS99}
\end{figure}

We now present our results. In Fig.~\ref{fig_ga_aor02_p0_DIS99}
we show $g_A = \Delta u - \Delta d$.
%There appears to be an $a^2$ gradient (in distinction to the other results)
%and a consequent `waggle' problem with the extrapolation.
There appears to be an $a^2$ gradient (in distinction to the other results).
It seems difficult to reach the phenomenological result. 
Perhaps this is due to a quenching effect. Also in 
this figure we show $(\delta u - \delta d)^{rgi}$.
It would seem that we have a milder $O(a^2)$ gradient and also that
$(\delta u - \delta d)^{rgi} \gsim g_A$. This might indicate
that a non-relativistic description of spin structure for the
`quenched' nucleon is reasonable. Note, however, that we are
well away from the non-relativistic limit of $5/3$.
To conclude, in the Table, we give our $n_f=0$ continuum results:
\begin{table}[h]
   \vspace*{-1.00cm}
   \begin{center}
      \begin{tabular}{||c|l||}
         \hline
         $g_A = \Delta u - \Delta d$     &  1.14(3)    \\
         $\Delta u$                      &  0.889(29)  $+\Delta q_{qldis}$ \\
         $\Delta d$                      &  -0.236(27) $+\Delta q_{qldis}$ \\
         \hline
         $(f_\rho^\bot / m_\rho)^{rgi}$  &  0.213(6)   \\
         \hline
         $(\delta u - \delta d)^{rgi}$   &  1.21(4)    \\
         $(\delta u)^{rgi}$              &  0.980(30)  \\
         $(\delta d)^{rgi}$              &  -0.234(17) \\
         \hline
      \end{tabular}
   \end{center}
   \vspace*{-1.15cm}
\label{table_results}
\end{table}

\section*{ACKNOWLEDGEMENTS}
\label{acknowledgement}

The numerical calculations were performed on the
Ape $QH2$ at DESY-Zeuthen and the Cray $T3E$ at ZIB, Berlin.

\end{document}